\documentstyle[aps,preprint]{revtex}

\begin{document}
\draft
\newcommand{\be}{\begin{equation}}

\newcommand{\ee}{\end{equation}}
\def\bq{\begin{eqnarray}}
\def\eq{\end{eqnarray}}
\def\t{\tilde}
\def\dex{\delta x}
\def\de{\delta}
\title{Action principle formulation for motion of extended bodies in General Relativity}

\author{Jeeva Anandan$^{\star ~ a}$, Naresh Dadhich$^{\dagger ~ b}$, 
Parampreet Singh$^{ \dagger ~ c}$}\address{${}^\star$ Department of Physics 
and Astronomy, University of South Carolina, \\Columbia, SC 29208, USA. 
\\
${}^\dagger$ Inter-University Centre for Astronomy and Astrophysics, 
\\Post Bag 4, Ganeshkhind, Pune~411~007, India.}

\maketitle
\vskip0.5cm
\begin{abstract}We present an  action principle formulation for the 
study of motion of an extended body in General Relativity in the limit of weak gravitational field. 
This gives the classical equations of motion for multipole 
moments of arbitrary order coupling to the gravitational field. 
In particular, a new force due to the octupole moment is obtained. 
The action also yields the gravitationally induced phase shifts in 
quantum interference experiments due to the coupling of all multipole moments. 
\end{abstract}
\vskip0.5cm
\vskip1cm

The study of motion of extended bodies possessing multipole moments in 
the gravitational field has a long history \cite{dixon1}. The starting 
point was the Einstein - Infeld - Hoffman \cite{eih} derivation of the 
geodesic equation for a point test particle from the gravitational field 
equation and the conservation law for stress energy tensor. The test 
particle approximation breaks down if body's extension in space is non 
negligible compared to the radius of curvature  of the background field 
and secondly when the back reaction due to the body on the background 
field is non ignorable. In this letter, we shall be concerned with the 
former aspect. This is particularly motivated by the fact that astrophysical 
bodies like planets and stars are extended and should in a realistic analysis 
be treated as such. The interaction of covariant generalization of 
Newtonian multipole moments with the gravitational field will be given 
by their coupling to Riemann curvature and its derivatives. This would 
appear as modification to the geodesic equation.

The modification to the 
geodesic for a spinning body is given by the Mathisson-Papapetrou 
equation \cite{mat,papa}, which may be extended to a particle with 
intrinsic spin \cite{jeeva4}. 
Subsequent to the treatment of spinning bodies, various authors have 
obtained the corrections up to the covariant generalization of Newtonian 
quadrupole moment \cite{taub,madore,dixon}.
 A comprehensive study of the problem including comparison of various 
approaches  and results is carried out by Dixon\cite{dixon}, but, to 
our knowledge, no one has obtained corrections to geodesic equation 
arising due to coupling of covariantized higher order Newtonian 
multipole moments with gravitational field. More importantly a 
procedure to derive equations of motion of extended bodies, with 
arbitrary multipole moments, through an action principle has not 
been obtained during the past 65 years in which the equations of motion 
in a gravitational field have been studied \cite{apb}. In the absence 
of a general principle to obtain the action made up of terms uniquely 
attributed to couplings of all multipole moments, such a task is very 
difficult. This is precisely what we wish to do in the following.

While the action is used in classical physics only as a tool to obtain 
the equations of motion, the action is directly observable in quantum 
physics as the  phase of the wavefunction. 
Therefore, the  phase shift produced by the coupling of multipole moments 
with the gravitational field can in principle be measured (action giving 
an algorithm to calculate various multipole phase shifts). Neutron 
interferometry provided the first instance where effect of the earth's 
gravitational field on the phase of the neutron wavefunction was 
observed \cite{cow}. 
Interesting gravitational analogs \cite{jeeva1} of the topological 
Aharanov-Bohm \cite{ab} and Aharanov-Casher \cite{ac,jeeva2} phase shifts 
have also been proposed. 

In this paper we  present a formalism which 
yields in a simple and elegant way the corrections to the geodesic 
equation up to all multipoles of the extended body. 
The  equations of motion for   multipoles simply follow from 
variation of our action. As a demonstration of our formalism we obtain 
for the first time corrections to the geodesic equation till the covariant 
generalization of Newtonian octupole moment. 
Moreover, our action gives the quantum phase shifts in interferometry 
due to the coupling of all multiple moments to the gravitational field. 
Our formulation of action principle for extended body may facilitate and 
prompt further investigations in the important and emerging area of 
interface between quantum and gravitational realms\cite{dvand}. 
Particularly, the ongoing experiments in atomic \cite{atomic}, 
molecular \cite{mol} and Bose-Einstein condensate 
interferometry \cite{bec} hold  promise for experimentally testing 
the new gravitational phase shifts that will be obtained in the present 
letter. 

We envision an extended rigid body as a thin world tube  in spacetime and its 
thickness is small compared to the scale over which curvature varies.
We would further assume that there are no external or internal forces acting
on this body apart from the gravitational field in which it is propagating.
In the thin world  tube we choose a reference world-line ($z^\mu$) 
having 4-velocity  
$u^\mu = dz^\mu/ds = (1, 0, 0, 0)$ and define multipole moments with respect 
to it on a spacelike hypersurface. The  multipole moments of order 2n are
 defined as \cite{dixon}
\be 
t^{\kappa_1 ... \kappa_n  \mu \nu} = \int \dex^{\kappa_1} ...{\dex^{\kappa_n}} \, \sqrt{-g} \, T^{\mu \nu} \, w^\alpha \, d \Sigma_\alpha \label{0}
\ee
where  $\de x^\mu = x^\mu - z^\mu$, $T^{\mu \nu}$ is the energy-momentum 
tensor, and the integration is 
over the spacelike hypersurface identified by the unit normal vector field $w^\alpha$. The above multipole moments are defined in a class of coordinate 
systems that are related by linear transformations in order that the 
expression (\ref{0}) is covariant. 
But once they are defined this way,  $ t^{\kappa_1 ... \kappa_n  \mu \nu}$ 
can be transformed to any arbitrary coordinate system as a tensor. 
All the relativistic equations in the present letter are covariant with 
respect to the above linear transformations, if not with respect to general 
coordinate transformations. 

Since what we do must be consistent with the 
Newtonian theory in the appropriate limit, we shall now establish the 
relation between the covariant multipole moments of order 2n and the 
covariant generalizations of the anti-symmetric spin tensor $S^{ij}$, 
the symmetric quadrupole moment tensor $I^{ij}$ and the symmetric octupole
 moment tensor $O^{ijk}$ in Newtonian gravity. We Write the Newtonian 
potential energy $U$ of the body with mass density $\rho(x)$ in terms 
of the potential $\phi(x)$ expanded in a Taylor series around the central 
world-line,
\be U =  \int \, \rho(x) \, \phi(x) \, d^3 x = m \, \phi(z) +  d^i \, \partial_i \phi |_z + \frac{1}{2} \, I^{ij} \, \partial_i \, \partial_j \, \phi |_z + \frac{1}{6} \, O^{ijk} \, \partial_i \, \partial_j \, \partial_k \, \phi |_z+ ... \label{eq:U}
\ee
where the mass $m = \int \rho(x) \, d^3 x$, the dipole moment $ d^i = \int \, \rho(x) \, \dex^i \, d^3 x, $ the quadrupole moment $ I^{ij} = \int \, \rho(x) \, \dex^i \, \dex^jd^3 x $ and the octupole moment $O^{ijk} = \int \, \rho(x) \, \dex^i \, \dex^j \, \dex^k d^3 x $, with  $\de x^i = x^i - z^i$. 
In view of eq.(\ref{0}), we identify $t^{i 00} = d^i, ~~ ~~ t^{ij00} = I^{ij}, ~~ ~~  t^{ijk 00} = O^{ijk}$. The Spin tensor(orbital angular momentum) 
in the Newtonian limit is defined as $S^{ij} = 2 \int \rho \, \de x^{[i} v^{j]} d^3 x $ where $v^i = d x^i/d t$. The Spin tensor then satisfies $d S^{ij}/d t = 2 \, p^{[i}  u^{j]} - 2 \int \rho \, \dex^{[i} \partial^{j]} \phi d^3 x$, where $u^i = d z^i/ d t$ and the momentum $p^i = \int \, \rho(x) \, v^i d^3 x$. 
Using Taylor expansion of the potential, and choosing $ z^i$ to be the 
center of mass so that $ d^i=0$, the spin propagation equation up to 
octupole term becomes
\be
\frac{d}{d t} S^{ij} = 2 \, p^{[i}  u^{j]} - 2 \, I^{k [i} \partial^{j]} \partial_k \phi|_z - O^{k r [i} \partial^{j]} \, \partial_k \partial_r \phi|_z \label{eq:spineq}
\ee
The covariantization of this spin tensor leads to 
\be \label{spin} 
S^{\mu \nu} =  t^{\mu \nu 0} - t^{\nu \mu 0} .
\label{eq:stensor}
\ee 
In the weak field limit, the metric is $g_{\mu\nu} = \eta_{\mu\nu} + h_{\mu\nu}$ with $h_{\mu \nu} \ll 1$. 
In the Newtonian limit $g_{00} = 1 + 2 \phi$ which implies $\phi = h_{00}/2$. 
Thus the dipole term in eq.(\ref{eq:U}) leads to the covariant form 
$t^{\mu \alpha \beta}\,  h_{\alpha \beta, \mu}|_z $, where $t^{\mu \alpha \beta}$ includes the spin tensor. We shall now choose the reference 
world line to be the center of mass so that $d^i = 0$. 
We consider only matter distributions for which \cite{dixon2}
\be\label{spin1}
t^{\mu \nu \alpha} = S^{\mu (\nu} u^{\alpha)} , 
\ee 
where$S^{\mu \nu}$ satisfies $S^{\mu \nu} u_\nu = 0$. 

The quadrupole term 
leads to $ \partial_i \, \partial_j \phi|_z \, I^{ij} = \frac{1}{2} \, \partial_i \, \partial_j h_{00}|_z I^{ij} =  - R_{0i0j}|_z \, I^{ij} $. 
We  thus have $ \partial_i \, \partial_j h_{00}|_z \, t^{ij00} = - 2 \, R_{0i0j}|_z \, I^{ij}$ which covariantizes to 
\be\label{quad1}
h_{\alpha \beta, \mu \nu}|_z \, t^{\mu \nu \alpha \beta} = - 2 \, R_{\alpha \mu \beta \nu}|_z \, I^{\mu \nu} \, u^{\alpha} \, u^{\beta}
\ee
where the covariant quadrupole tensor, $I^{\mu \nu} = I^{\nu \mu}$. 
Similarly, the octupole term leads to
\be \label{oct1}
h_{\alpha \beta, \mu \nu \sigma}|_z \, t^{\mu \nu \sigma  \alpha \beta} = - 2 \, R_{\alpha \mu \beta \nu, \sigma }|_z \, O^{\mu \nu \sigma} \, u^{\alpha} \, u^{\beta} ~~
\ee
where $O^{\mu \nu \sigma }$ is the fully symmetric covariant octupole tensor. 
Eqs.(\ref{spin1}), (\ref{quad1}) and (\ref{oct1}) are the key relations 
which would unambiguously provide the connection between covariant multipole 
moments with their Newtonian analogs. 

In the Newtonian approximation the 
phase shift in quantum mechanical interference due to the gravitational field 
is\be
\Phi = -{ 1\over \hbar} \int U dt =-{ 1\over \hbar}(\int m \, \phi dt +  \int d^i \, \partial_i \phi dt+\int \frac{1}{2} \, I^{ij} \, \partial_i \, \partial_j \, \phi dt + \int \frac{1}{6} \, O^{ijk} \, \partial_i \, \partial_j \, \partial_k \, \phi dt+ ... )\label{phase}
\ee
where $U$ is given by the expansion in using eq.(\ref{eq:U}). 
The first term of eq.(\ref{phase}) corresponds to the phase shift observed in 
the COW experiment \cite{cow}, and the subsequent dipole and quadrupole terms 
are corrections to it. The higher order multipole contributions to the phase 
shift may also be 
obtained from this expansion. In General Relativity this phase shift is 
obtained by letting the path ordered operator resulting from the
 covariant generalization of this Newtonian phase shift act on the initial 
wavefunction. Using eqs.(\ref{eq:U}, \ref{spin1}, \ref{quad1}, \ref{oct1}) 
and noting that in the linear field limit  $\omega^a{}_{b \, \beta} S^b{}_a = h_{\alpha \beta, \mu} S^{\alpha \mu}$, where  $\omega^a{}_{b \, \beta}$ 
are Ricci rotation coefficients,this path ordered operator is given by
\be
g = {\cal P} exp \bigg[-\frac{i}{\hbar} \int (- ~ m  + \frac{1}{2} \, \omega^a{}_{b \beta} S_a{}^b \, u^\beta - \frac{1}{2} \, R_{\alpha \mu \beta \nu}I^{\alpha \beta}\, u^\mu u^\nu - \frac{1}{6} \, \int \,    R_{\alpha \mu \beta \nu; \rho} \, O^{\alpha \beta \rho} \, u^\mu \, u^\nu +...) ds \bigg] ~~. \label{eq:pathop}
\ee 
In the special case of $I^{\alpha \beta}$ and $O^{ijk}$ being zero, this 
result is in agreement with the gravitational phase shift for intrinsic 
spin \cite{jeeva4}. 

We shall now obtain this expression in the weak field 
limit of General Relativity starting from the action principle. Choose a 
coordinate system such that along the reference world-line 
$g_{\mu\nu} = \eta_{\mu\nu}$. We would here like to recall that the extended
body under consideration is rigid and subject to only external gravitational 
force and no other forces (external or internal). In that case the monopole moment corresponds to the 
mass ($m$) of the body 
with whole matter concentrated on the reference world-line and ignoring the
 back reaction on the background gravitational field we write $m = p_\alpha \, u^\alpha$. The  higher order multipole moments are defined by considering 
the metric perturbations as one moves away from the reference 
world-line $z^\mu$. For simplicity we restrict perturbations of the 
metric to the first order \cite{secondorder}, and later covariantize the equation of motion
 by changing ordinary derivative to covariant derivative. The action
\be{\cal S} = \int \sqrt{-g} \, {\cal L} \, d^4 x = \int \sqrt{-g} \, {\cal L}|_{g_{\mu \nu} = \eta_{\mu \nu}} \, d^4 x + \left(\int \sqrt{-g} \, {\cal L} \, d^4 x - \int \sqrt{-g} \, {\cal L}|_{g_{\mu \nu} = \eta_{\mu \nu}} \, d^4 x \right) .
\ee
The first term is the kinetic energy term,  $\int \sqrt{-g} \, {\cal L}|_{g_{\mu \nu} = \eta_{\mu \nu}} \, d^4 x  = - \,\int p_{\alpha} \, u^{\alpha} \, d s $. So,
\be
{\cal S} = - \,  \int p_{\alpha} \, u^{\alpha} \, d s + \int \de g_{\mu \nu} \, \frac{\de}{\de g_{\mu \nu}} (\sqrt{-g} {\cal L}) \, d^4 x + ... = - \,  \int p_{\alpha} \, u^{\alpha} \, d s + \frac{1}{2} \, \int \de g_{\mu \nu} \, \sqrt{-g} \, T^{\mu \nu} \, d^4 x + ... \label{eq:var}
\ee
Note $\de g_{\mu \nu} = g_{\mu \nu} - \eta_{\mu \nu}=h_{\mu \nu}$ and 
hence it can be written as $\de g_{\mu \nu} =  h_{\mu \nu, \sigma}|_z \, \de x^{\sigma} + \frac{1}{2} \,h_{\mu\nu, \sigma \rho}|_z \, \de x^{\sigma} \, \de x^{\rho} + ... $where we have  $h_{\mu \nu}(z) = 0$. In the present weak 
field limit, we neglect all terms that are quadratic or higher order in 
metric perturbations. Substituting for $\de g_{\mu \nu}$ in  
eq.(\ref{eq:var}) and using eqs.(\ref{0},\ref{spin1},\ref{quad1},\ref{oct1}), 
we finally 
obtain
\be \label{eq:action}
{\cal S}  =  - \,  \int p_{\alpha} \, u^{\alpha} \, d s + \frac{1}{2} \, \int h_{\alpha \beta, \mu} \, S^{\mu \alpha} \, u^{\beta} \, d s - \frac{1}{2} \,\int \,  R_{\alpha \mu \beta \nu} \, I^{\alpha \beta} \, u^{\mu} \, u^{\nu} \, d s - \frac{1}{6} \, \int \,  R_{\alpha \mu \beta \nu, \rho} \, O^{\alpha \beta \rho} \, u^\mu \, u^\nu \, d s
\ee up to octupole (all derivatives are evaluated on $z^\mu$). 
In the linear field limit, since $h_{\alpha \beta, \mu} \, S^{\mu \alpha} = \omega^a{}_{b \beta} S_a{}^b$, hence the accumulation of infinitesimal phases 
arising from eq.(\ref{eq:action}) is the same as what is obtained from 
the path ordered operator eq.(\ref{eq:pathop}) in this limit. 

We now 
obtain the equation of motion by extremizing the action $(\delta S = 0)$ 
and requiring that coordinate variations vanish at end points of the path. 
This leads  to the equation of 
motion
\bq
\frac{d p_{\sigma}}{ds} &=& \nonumber  R_{\sigma \lambda \mu \nu} \, \left(\frac{1}{2} \, u^\lambda \, S^{\mu \nu}+ \frac{d}{ds} \left(I^{\lambda \mu} \, u^\nu \right) \right) + \frac{1}{2} \,  h_{\nu \sigma, \mu} \, \frac{d}{d s} \, S^{\mu \nu} \\&~~ & -  R_{\sigma \lambda \mu \nu, \rho} \, u^{\mu} \, I^{\nu(\rho} \, u^{\lambda)} + \frac{1}{3} \, R_{\alpha \mu \beta \sigma, \nu \rho} \, \, u^\alpha \, O^{\mu \rho (\nu} \, u^{\beta)} - \frac{1}{3} \, R_{\alpha \mu \sigma \nu, \rho} \, \frac{d}{d s} \left(O^{\mu \nu \rho} \, u^\alpha \right)\label{noncovariant}
\eq
In a coordinate system in 
which $h_{\alpha\beta,\mu}=0$, (\ref{noncovariant}) gives 
\bq
\frac{D p_{\sigma}}{Ds} &=& \nonumber  R_{\sigma \lambda \mu \nu} \, \left(\frac{1}{2} \, u^\lambda \, S^{\mu \nu}+ \frac{D}{Ds} \left(I^{\lambda \mu} \, u^\nu \right) \right) -  u^{\mu} \, I^{\nu(\rho} \, u^{\lambda)} \, \nabla_\rho  R_{\sigma \lambda \mu \nu} \\& ~~ & + \frac{1}{3} \, \nabla_\nu \nabla_\rho R_{\alpha \mu \beta \sigma} \, \, u^\alpha \, O^{\mu \rho (\nu} \, u^{\beta)} - \frac{1}{3} \, \nabla_\rho \, R_{\alpha \mu \sigma \nu} \, \frac{D}{D s} \left(O^{\mu \nu \rho} \, u^\alpha \right)\label{covariant} .
\eq 
Eq.(\ref{covariant}) is 
generally covariant and thus is valid in every coordinate system.
Since the dipole moment couples to Riemann curvature, it is expected that 
quadrupole should couple to its first derivative and octupole to its 
second derivative. However there is a  coupling of quadrupole with Riemann 
curvature in the second term and a  coupling of octupole with first 
derivative of Riemann curvature in the last term which  suggests that 
$p_\alpha$ should be suitably redefined \cite{dixon3} as, $p^*_{\sigma} = p_\sigma - R_{\sigma \lambda \mu \nu} \, I^{\lambda \mu} \, u^\nu  - \frac{1}{3} \nabla_\rho R_{\sigma \mu \nu \alpha} O^{\mu \nu \rho} u^\alpha $, 
which would then yield the expected form
\be
\frac{D p^*_{\sigma}}{Ds} =  \frac{1}{2} \, R_{\sigma \lambda \mu \nu} \, u^{\lambda} \, S^{\mu \nu} + \frac{1}{2} \, \nabla_{\sigma} R_{\alpha \mu \beta \nu} \, u^{[\mu} I^{\alpha ][\beta} u^{\nu ]} + \frac{1}{6} \, \nabla_\rho \nabla_\sigma \, R_{\alpha \mu \beta \nu} \, u^{[\alpha} O^{\mu] \rho [\nu} u^{\beta]} ~~.\label{4}
\ee
This is the equation of motion for a body possessing dipole, quadrupole 
and octupole moments in a gravitational field. 
The spin propagation equation, eq.(\ref{eq:spineq}), can also be 
covariantly generalized to
\be
\frac{D}{D s} S^{\alpha \beta} = 2 \, p^{[\alpha} \, u^{\beta]} - 2 \, R^{[\alpha }{}_{\mu \nu \sigma} \, u^{[\mu} I^{\beta]][\nu} u^{\sigma]} - R^{[\alpha}{}_{\mu \nu \sigma; \rho} \, u^{[\mu} \, O^{\beta]] \rho [\nu} \, u^{\sigma]} \ee
which on the  redefinition of momentum vector form $p_\alpha$ to $p^*_\alpha$modifies as follows:
\bq
\frac{D}{Ds} S^{\alpha \beta} &=& \nonumber 2 \, p^{* [\alpha} \, u^{\beta]} + 2 \left( R^{\alpha}{}_{\mu \nu \sigma} \,  u^{[\beta} I^{\mu ][\nu} u^{\sigma]} - R^{\beta}{}_{\mu \nu \sigma} \,  u^{[\alpha} I^{\mu ][\nu} u^{\sigma]}  \right)\\&~~ & - 2 \, R^{[\alpha }{}_{\mu \nu \sigma} \, u^{[\mu} I^{\beta]][\nu} u^{\sigma]} - \frac{5}{3} \, R^{[\alpha}{}_{\mu \nu \sigma; \rho} \, u^{[\mu} \, O^{\beta]] \rho [\nu} \, u^{\sigma]}~~.
\eq

To simplify the notation we would further define  $J^{\mu \alpha \beta \nu} := - 3 \, u^{[\mu} I^{\alpha ][\beta} u^{\nu ]} $ and $G^{\mu \alpha \beta \sigma \nu} := -  u^{[\mu} O^{\alpha ]\beta [ \sigma} u^{\nu ]} $ and then finally 
obtain
\be
\frac{D p^*_{\sigma}}{Ds} =  \frac{1}{2} \, R_{\sigma \lambda \mu \nu} \, u^{\lambda} \, S^{\mu \nu} + \frac{1}{6} \, J^{\mu \alpha \beta \nu} \, \nabla_{\sigma} R_{\mu \alpha \beta \nu} + \frac{1}{6} \, \nabla_\rho \nabla_\sigma \, R_{\alpha \mu \beta \nu} \, G^{\alpha \mu \rho \nu \beta} ~~.\ee\be\label{7}\frac{D}{Ds} S^{\alpha \beta} = 2 \, p^{* [\alpha} \, u^{\beta]} - \frac{4}{3} \, R^{[\alpha }{}_{\mu \nu \sigma} \, J^{\beta] \mu \nu \sigma} - \frac{5}{3} \,R^{[\alpha }{}_{\mu \nu \sigma ; \rho} \, G^{\beta] \mu \rho \nu \sigma} ~~.
\ee
Thus we have obtained for the first time the correction to propagation 
equations till the coupling of covariant generalization of Newtonian 
octupole moment with the background gravitational field. These equations 
agree with the earlier results obtained till quadrupole \cite{dixon} but 
the force due to the octupole moment is a new result. Our procedure can be 
easily generalized  easily extended to obtain further corrections from due 
to all higher multipole moments.

This is a very simple and elegant method of 
deriving the equation of motion for an extended body incorporating coupling
 of multipole moments of arbitrary order with the gravitational field.
 And above all we have found the action for such a body which we have 
also used  to compute  the gravitational  phase shifts in its quantum
 wavefunction  due to the multipoles. We hope that the new gravitationally 
induced phase shifts would be measured in future interferometry experiments 
based on atomic, molecular and Bose-Einstein condensates \cite{atomic,mol,bec}. Our novel algorithm also yields  modifications to the geodesic equation 
which, with present day high precision astronomical observations yielding 
multipole moments of planetary or stellar bodies,can be applied to obtain 
aberrations in their orbits. It would be interesting to relax the other 
aspect of test body character; i.e. non namely
 ignorability of back reaction, and then study the motion in full 
generality. 

Acknowledgements: We thank J. Ehlers and C. W. Misner for 
informative discussions. PS thanks Council for Scientific and Industrial 
Research for grant number: 2-34/98(ii)E.U-II.


\begin{references}
\bibitem{secondorder} If the size of the body is $L$ and the radius
of curvature of the background spacetime is $R$, restricting to the 
first order would imply that we
can go up to octupole. This is because the metric perturbation goes as
$L^2/R^2$ while the expansion in multipole coupling goes as $L/R$. That
means octupole would go as $L^3/R^3$ which would be fine. However, if one
wishes to go to higher multipole expansion, one would have to include the 
second order term in the metric perturbation.
\end{references}
\end{document}